\documentclass[journal]{IEEEtran}
\usepackage{graphicx} 
\usepackage{cite}
\usepackage{algorithm}
\usepackage{algorithmic}
\newcommand{\INDSTATE}[1][1]{\STATE\hspace{#1\algorithmicindent}}
\usepackage[table]{xcolor}
\usepackage{pifont}
\usepackage{wasysym}
\usepackage[normalem]{ulem}

\begin{document}

\title{Analog Isolated Multilevel Quantizer for Voltage Sensing while Maintaining Galvanic Isolation}

\author{Peter~Weber,~\IEEEmembership{Member,~IEEE,}
        and Antonia~Papandreou-Suppappola,~\IEEEmembership{Fellow,~IEEE}
\thanks{P. Weber and A. Papandreou-Suppappola are with the Department
of Electrical and Computer Engineering, Arizona State University, Tempe,
AZ, 85281 USA e-mail: (see peter.c.weber@asu.edu).}
\thanks{Manuscript submitted February 28, 2024.}
}

\markboth{}%
{Shell \MakeLowercase{\textit{et al.}}: AIMQ}

\maketitle

\begin{abstract}
A low-power, compact device for performing measurements in electrical systems with isolated voltage domains is proposed. Isolated measurements are required in numerous applications. For instance, a measurement of the bus voltage for a system with a high supply voltage and lower isolated local voltage level may be needed for system health monitoring and control. Such a requirement may necessitate the use of isolation amplifiers to provide voltage telemetry for the local system. Isolation amplifiers require dual galvanically isolated supplies and use magnetic, capacitive, or optical barriers between primary and secondary
sides. Producing this supplemental voltage requires an extra voltage converter, which consumes power and generates electromagnetic interference which must, in turn, be filtered. Complex designs incorporating feedback are needed to achieve
linear response. The proposed Analog Isolated Multilevel Quantizer (AIMQ) addresses these issues by monitoring the primary-side signal and communicating the results to the secondary side using a novel scheme involving Zener diodes,
optocouplers, transistors, one-hot coding, and discrete outputs. The result is a low power isolated transducer that can in principle be extended to an arbitrary bit depth.
\end{abstract}


\section{Introduction}
\IEEEPARstart{I}{n} systems
that supply dedicated
voltages to various loads from a power source operating at a higher voltage, there are  many circumstances  in which an isolation barrier is required to prevent direct current (DC) from flowing between the source and the load. Applications such as motor drives \cite{MS-2488}, biomedical sensors \cite{Zhou15}, Industrial process control \cite{FDI}, and power supplies \cite{ADuM4190}, require measurements that can be communicated across isolated voltage domains with different common-mode voltage levels. A widely used approach to producing this voltage telemetry is an isolation amplifier \cite{Heb77,Irv09}, with which it is possible to obtain the necessary low-level measurements.  

\section{Isolation Amplifiers}
\begin{figure}
\centering
\includegraphics[scale=.4]{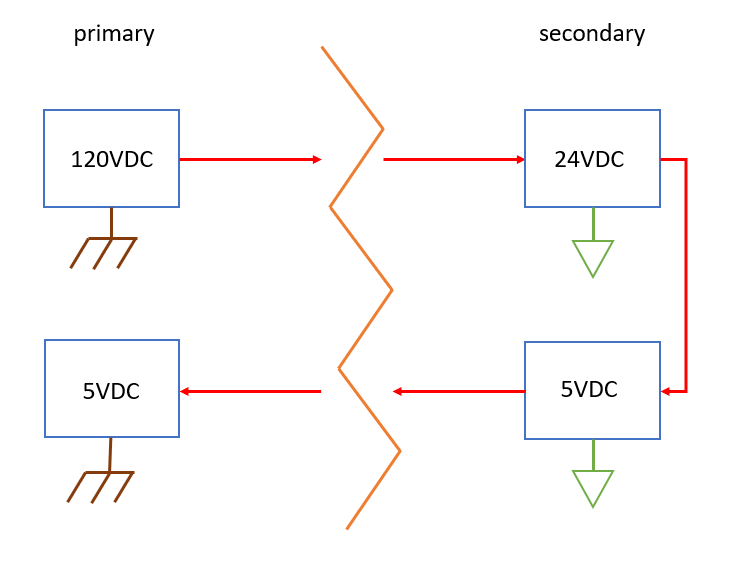}
\caption{Convoluted power scheme required for isolation amplifier primary-side power. The jagged line represents the magnetic barrier.}
\label{convoluted}
\end{figure}

Isolation amplifiers require dual galvanically isolated supplies and use magnetic \cite{AD202}, capacitive \cite{AMC-1211Q1}, or optical \cite{Ava15} barriers between primary and secondary sides.  The additional primary-side converter consumes power and generates electromagnetic interference that must be suppressed in order to meet industrial or military Electromagnetic Emissions and Compatibility (EMI/EMC) standards such as IEC EN 61000-3-2 \cite{Std61000}, IEEE Std 519-2014 \cite{Std519}, and MIL-STD-461G \cite{Std461}. Furthermore, these devices often  employ convoluted analog-to-digital converter (ADC) encoding methods, followed by immediate digital-to-analog converter (DAC) signal reconstruction, in order to generate an alternating current (AC) signal that can cross the galvanic barrier. An example of this approach is shown in Figure \ref{convoluted}.

\subsection{Analog to Digital to Analog Conversion}
In comparison to a general purpose operational amplifier, the need for isolation results in additional complexities to achieve equivalent output linearity and bandwidth. For instance, magnetically-coupled devices must generate a time varying signal with sufficient $\frac{d\phi}{dt}$ to induce the output voltage in accordance with Faraday's Law of Induction:
\begin{equation}
    V=-\frac{d\phi}{dt}=-\frac{d}{dt}\int_A \mathbf{B}(t)\cdot d\mathbf{A}
\end{equation}
where, according to Ampere's Law:
\begin{equation}
    \oint \mathbf{B} \cdot d\mathbf{l} = \mu I_{enc}
\end{equation}
and $\mu$ is the magnetic permeability of the core material providing the flux linkage between the primary and secondary domains.

This can be achieved by converting the analog signal to its digital representation via some form of Pulse Code Modulation (PCM). The A-to-D conversion process generally begins with a Low-Pass Filter (LPF) for antialiasing, followed by a clocked sample and hold circuit converting the continuous time analog signal $x_a(t)$ to a discrete time signal $x_a[n]$, which is then input to a quantizer. The quantizer output is then encoded, forming a discrete time digital representation $x_d[n]$ of the input signal \cite{Gag75,Sko82,Che12}. 

\begin{figure}
\centering
\includegraphics[scale=.32]{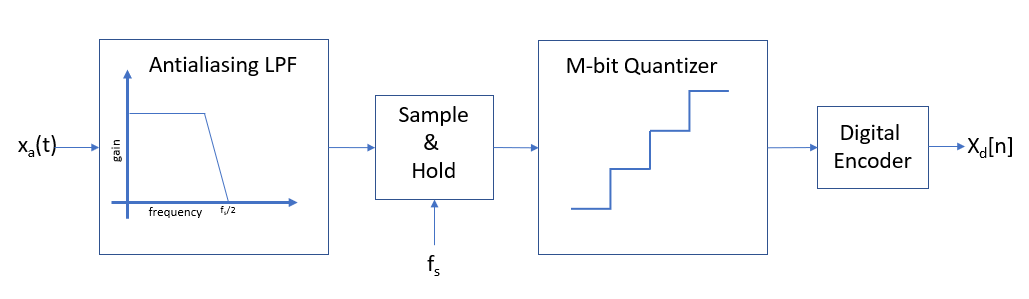}
\caption{Block Diagram of Analog to Digital Converter}
\label{ADC_blockdiagram}
\end{figure}

\subsection{Sigma-Delta Modulation} 
The quantizer shown in Figure \ref{ADC_blockdiagram} generally consists of a set of comparators with defined thresholds. An $M$-bit realization of such a quantizer would therefore require $M$ comparators and $M$ reference voltages. Delta modulation eliminates these additional levels by simply comparing the current value with the previous one and recording the difference by adding or removing a fixed quantity $m\Delta$, where $m=1,...,M$ is the number of quantization levels, and $\Delta$ is the difference between successive levels. Sigma-Delta ($\Sigma\Delta$) modulation \cite{Sigma} improves linearity through additional feedback to shape the quantization noise. This method achieves quantization of the input signal by means of a comparator with its noninverting input tied to the analog signal of interest and its inverting terminal tied to its filtered output. Since the desired output of an isolation amplifier is often an analog signal, this modulated output then must be demodulated through an additional DAC. This approach is frequently used in both optical and capacitive isolation amplifiers \cite{AMC-1211Q1, Ava15}.

\subsection{EMI Ramifications of Isolation Amplifiers}
The need for additional power conversion for the primary side of the isolation amplifier adds more switching noise, a significant source of electromagnetic interference (EMI) \cite{TrincheroRiccardo2015EPoS}. In most cases, this means that in addition to the extra isolated DC/DC converter, an EMI filter such as a passive second-order inductor-capacitor (LC) filter is likely to be needed. Along with the requisite converter output filter and any potential minimum load current resistor recommended by the converter manufacturer for proper voltage regulation \cite{CRE1s0505}, the increased parts count can become significant.

\subsection{Internal Primary-Side Voltage Generation}
Modern devices are beginning to appear in the marketplace advertising internal generation of the required primary-side power \cite{AMC3302}, and also performing ADC via $\Sigma\Delta$ modulation to generate the bitstream that crosses the isolation barrier, followed by Digital to Analog conversion on the secondary side, and, in the cited example, including a 4th-order analog output filter. Adding to this complexity, the design includes the internal isolated DC/DC converter. With continued miniaturization and power reductions in CMOS fabrication processes, this approach appears unchallenged in the marketplace. 

\section{An Alternative for Isolated Measurements}
We propose a method of performing isolated measurements that greatly simplifies the process, reduces parts count, decreases EMI, and gives up little in the way of measurement accuracy. We call this device the Analog Isolated Multilevel Quantizer (AIMQ) and our method is summarized in Algorithm \ref{AIMQ algorithm}.

\begin{algorithm}
\caption{AIMQ Algorithm}
\label{AIMQ algorithm}
\begin{algorithmic}[1]
\STATE Monitor primary side input voltage $V_{pri}$ via Zener diode thresholding
\INDSTATE  Zener diode stacks with $V_Z \leq V_{pri}$ will conduct
\INDSTATE  Determine number $N$ of Zener stacks for desired measurement resolution
\INDSTATE  $V_{Z,min}=V_{Z_1}, ..., V_{Z_N}= V_{Z,max}\leq V_{pri,max}$
\INDSTATE \hspace{1cm} \ding{43} (where $V_{pri,min}\leq V_{Z,min}$)
\STATE Implement One-hot encoding
\INDSTATE  Disable all Zener stacks with $V_Z < V_{Z,hi}$ 
\INDSTATE \hspace{1cm} \ding{43} (where $V_{Z,hi}$ is the highest voltage Zener  stack in conduction)
\INDSTATE  Generate "Level$_{n-1}$turnoff" signal
\INDSTATE  Route "Level$_{n-1}$turnoff" to $V_{Z_{n-1}}$ pull-down transistor
\INDSTATE  Route pull-down collector to $n-1$ base resistor of series-pass turn-off transistor
\INDSTATE  diode OR turnoff signal to all lower Zener stack series-pass transistors
\STATE Transmit $V_{Z,hi}$ stack current $I_{Z,hi}$ across optocoupler
\INDSTATE  Scale optocoupler output voltages by scale factor $\gamma$
\INDSTATE \hspace{1cm} \ding{43} $\gamma = V_Z/V_{ADC_{max}}$ where $V_{ADC_{max}}$ is ADC 
input maximum voltage
\INDSTATE  $V_{out_{max}}=V_{ADC_{max}}$, $V_{out_{n}}=V_{ADC_{max}}\frac{n}{N}$
\end{algorithmic}
\end{algorithm}

The proposed device uses a set of discrete outputs that are enabled only when the primary-side voltage exceeds various threshold levels. The device has a similar function to existing quantizers such as ADCs \cite{GrayR.M.1998Q}, but it uses an analog output scheme inspired by the principle of one-hot encoding \cite{GuBonwoo2021ERLM} to minimize power. 

\subsection{One-Hot Encoding}
In a one-hot encoding scheme, a quantity is encoded using multiple bits, only one of which is enabled at any time. In most cases this approach trades resource efficiency for error immunity \cite{GuBonwoo2021ERLM} or for numerical encoding of categorical data \cite{KunanbayevKassymzhomart2021CE}, but here it is used to minimize primary-side measurement current, and thereby \textit{improves} efficiency. For instance, a 3-bit quantity can take on $2^3=8$ distinct values, and would require an 8-bit word to represent it in a one-hot scheme. Table \ref{one-hot table} shows the scheme. Note that there can be some ambiguity regarding how to encode a zero value, and we choose to use all zeroes, allowing $2^3+1=9$ possible values from the 8-bit one-hot scheme. 

\begin{figure}
\centering
\includegraphics[scale=.35]{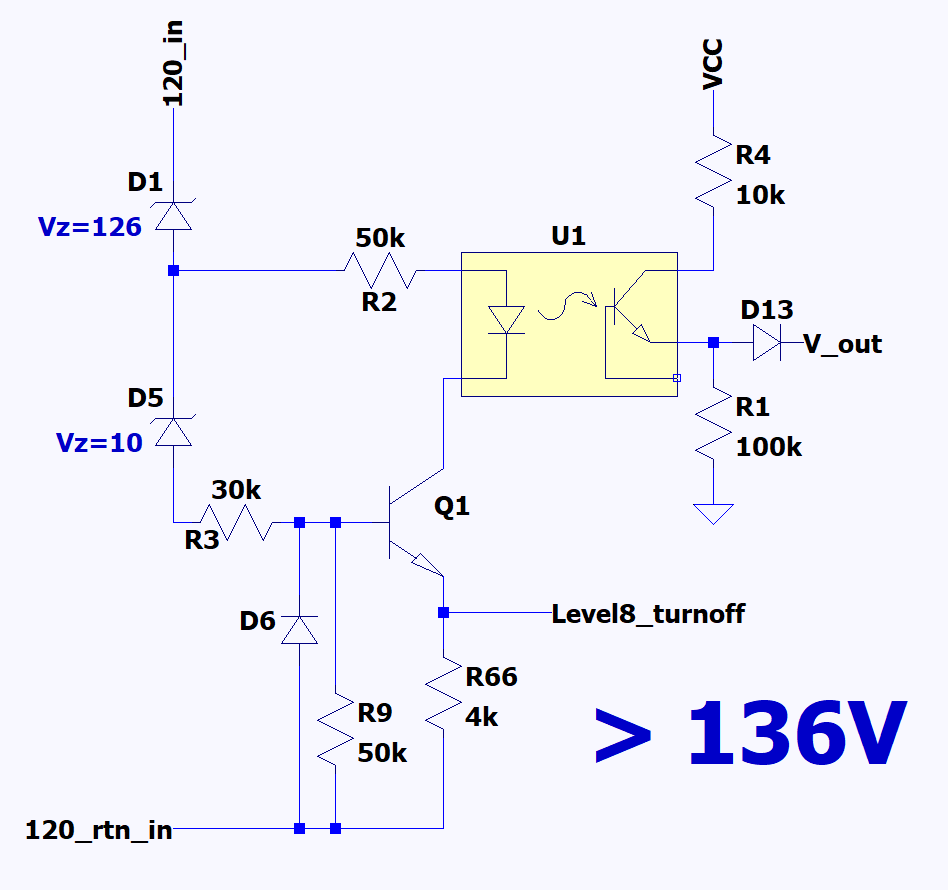}
\caption{Functional Unit of the AIMQ. This circuit will output a scaled secondary-side signal $V_{out}$ when the primary-side voltage exceeds 136 V.}
\label{AIMQ functional unit}
\end{figure}

There are multiple secondary-side outputs, each of which is powered by the same secondary power, but divided down to different voltage levels according to the input level they are intended to represent. This process effectively abstracts away the details of the analog to digital to analog conversion by performing all of these steps in the analog realm. The outputs
all connect to a single node via a diode ORing scheme. The input is powered by the primary-side voltage, and works as follows:

Assume a nominal primary bus voltage $V_{pri}$, and a secondary-side circuit that runs on isolated $V_{sec}$. Note that in Figure \ref{AIMQ functional unit}, $V_{pri}= 120_{in}$ and $V_{sec}=V_{CC}$. The secondary electronics require a measurement of the bus voltage, either for control or telemetry purposes. Assume further that we require n-bits of precision in the voltage telemetry. In fact, the maximum useful bit-depth is determined by the choice of downstream ADC, and any additional analog precision is wasted. Furthermore, in some cases, an ostensibly linear transducer produces a transfer characteristic that is better fitted to a polynomial of order $m\geq3$ (observed), in which case there is some ambiguity in the input value corresponding to certain outputs, and the supposed precision of the analog signal is illusory. 

\begin{table}
\begin{center}
\begin{tabular}{llll}
\textbf{\uline{Binary}} & \textbf{\uline{One-hot}} &  &   \\
000                     & 00000000                 &  &   \\
001                     & 00000001                 &  &   \\
010                     & 00000010                 &  &   \\
011                     & 00000100                 &  &   \\
100                     & 00001000                 &  &   \\
101                     & 00010000                 &  &   \\
110                     & 00100000                 &  &   \\
111                     & 01000000                 &  &   \\
1000                    & 10000000                 &  &  
\end{tabular}
\end{center}
\caption{Binary to one-hot conversion}
\label{one-hot table}
\end{table}

\subsection{The AIMQ Functional Unit}
The functional unit of the Analog Isolated Multilevel Quantizer is shown in Figure \ref{AIMQ functional unit}. The complete AIMQ has $2^n$ such optocouplers in parallel, each of which is only allowed to conduct primary-side current when its Zener diode ladder's voltage level is exceeded, and where $n$ is the bit-depth of the signal. The diode ladder sets the threshold for its optocoupler's output to be enabled. We characterize this as a ladder (a stack of diodes) to simplify the parts procurement process, but it is equally valid to use specific individual Zener diodes to generate the various voltage thresholds. The significance of the Zener diode is its highly nonlinear IV characteristic, which implies a predictable turn-on threshold for each channel. The tuned breakdown voltage is a balance between the tunnel and avalanche breakdown mechanisms for current flow in a reverse-biased diode. 

However, this also implies that measures must be taken to limit the breakdown current, otherwise the diodes will be be damaged. The series npn transistor Q1 shown in Figure \ref{AIMQ functional unit} serves this function. The collector current $I_C$ for Q1 must be equal to the optocoupler U1 input LED forward current $I_f$ ($I_C=I_f$), and this equates to a Q1 base current:
\begin{equation}
    I_B=\frac{I_C}{\beta}
\end{equation}
where $\beta \approx 100$ is the current gain of Q1. The Zener diode ladder current is split between the input to the optocoupler and the Q1 base. The voltage divider formed by R3 and R9 ensure the the Q1 base voltage is appropriately scaled, and the switching diode D6 between the 120 VDC return and the base of Q1 allows fast turn-off. 

Each output is scaled to represent the input voltage corresponding to its conduction threshold. Furthermore, the inputs of the parallel optocouplers are daisy-chained such that if a higher level threshold is enabled, all lower levels inputs are disabled by means of a high-side series pass transistor. In this way, input power is minimized by using an analog equivalent to a digital one-hot encoding scheme.

\subsection{Minimizing Primary-Side Power Consumption}
When performing measurements for telemetry, it is important to be efficient. Particularly when primary-side voltages are large in magnitude, the current consumed by any voltage dividers, for instance, should be kept to a minimum. In the AIMQ, this current minimization is achieved in concert with the one-hot encoding, by turning off all lower-voltage Zener diode stacks through the use of a turnoff signal from higher level channels. This is shown in Figure \ref{AIMQ functional unit} as the "Level8\_turnoff" signal seen near the bottom of the schematic. 

\begin{figure}
\centering
\includegraphics[scale=.38]{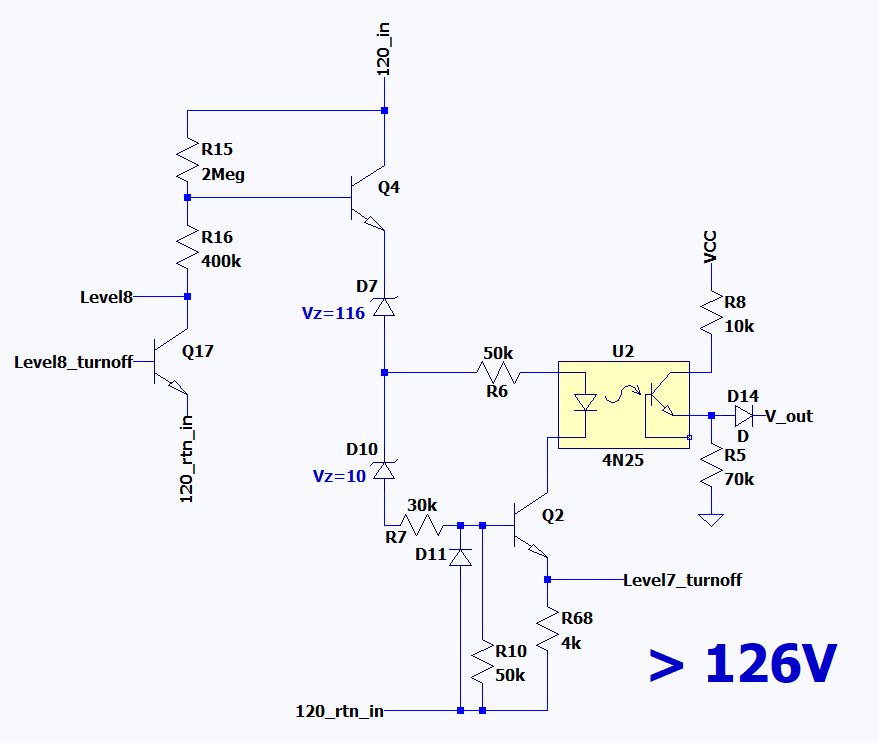}
\caption{Turnoff mechanism to disable all lower voltage channels and implement the one-hot encoding scheme. The Level 7 channel is shown.}
\label{turnoff mechanism}
\end{figure}

This turnoff signal is fed into a cascade of ORing diodes to all lower channels. When the primary voltage exceeds 136 V, the Level 8 channel's Zener stack begins to conduct, turning on the npn transistor Q1, and pulling the Q1 emitter up to a voltage $V=I_f \cdot R66$, determined by the optocoupler's input LED forward current $I_f$ and the pull-down resistor R66 in Figure \ref{AIMQ functional unit}. This voltage is the "Level8\_turnoff" signal, and is fed to the Channel 7 pull-down transistor Q17 shown in the upper right of Figure \ref{turnoff mechanism}.

\begin{figure}
\centering
\includegraphics[scale=.425]{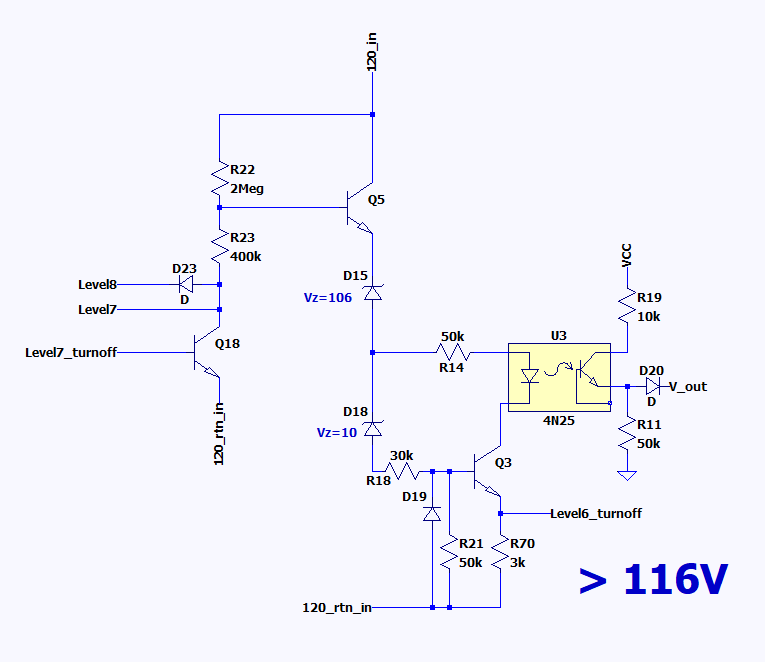}
\caption{The Level 6 channel in this case corresponds to an input voltage greater than 116 V due to the breakdown voltages of D15 and D18.}
\label{116channel}
\end{figure}

When Q17 turns on, this robs the series pass transistor Q4 of its base current, and the Zener stack below Q4 is disabled. Figure \ref{116channel} in turn shows the next lower channel, and there is a diode connecting the collector of Q18 to the collector of Q17 (labeled "Level8" in Figure \ref{turnoff mechanism}), which therefore pulls the node marked "Level 7" (shown in Figure \ref{116channel}) low through the diode D23, in turn disabling the 116V Zener stack. Each subsequent lower level follows the same scheme, ensuring that all lower levels are disabled by the highest active voltage level.

The Zener diode thresholding approach enables the quantizer to be employed in systems with large voltage potential differences. Further reductions in power can be achieved by using an optocoupler with maximum current transfer ratio (CTR) \cite{WangZhongqiang2009Troc} and minimal forward current. The levels should be chosen to prevent any Zener ladder from conducting large currents, as the diode IV curve is highly nonlinear above the breakdown voltage \cite{15KESeries}. Alternatively, the high-side series-pass transistors can be used to limit the Zener current. Power consumption can also be potentially reduced with optocoupler efficiency gains, and signal fidelity can be improved by increasing the number of output levels.

\subsection{A More Efficient Telemetry Scaling Scheme}
\label{usable_scale}
It is not unusual to scale a voltage telemetry signal in an inefficient way, such as a scale whereby a primary-side voltage range of 0 to 140 V corresponds to a telemetry signal of 0 to 5 V. In the case in which all downstream secondary-side logic circuits lose power when the primary voltage drops below 90 V due to the undervoltage lockout \cite{uvlo} of the DC/DC converter providing secondary isolated power, then  $>64\%$ (90V/140V) of the telemetry scale is wasted. Recognizing this, we can design the AIMQ to use a minimum channel voltage that is similar to the undervoltage lockout (UVLO) voltage of the converter producing the secondary-side voltage. 

Each level's output voltage is scaled by means of the value of the pull-down resistor at the optocoupler output. Although the circuit looks like a resistive voltage divider, the output voltage is in fact a function of the optocoupler's current tranfer ratio (CTR) multiplied by the pull-down resistor value (R1, R5, and R11 in Figures \ref{AIMQ functional unit}, \ref{turnoff mechanism}, and \ref{116channel}, respectively. In our simulations, we used a minimum channel voltage of 66 V; in practice this level will be determined by the minimum operating voltage of the overall system. 

\section{SPICE Simulations of the AIMQ}
\label{AIMQ spice}
A random noisy bus voltage was generated to produce an input signal that rises or falls by noise $V_{noise}[nT_s]$ drawn from a uniform distribution between 0 and 1  every 10 ms, or does not change. The resulting bus voltage is:
\begin{equation}
 V_{bus}[nT_s]= V_{bus}[(n-1)T_s]+V_{noise}[nT_s]
\end{equation}
where $V_{noise}[nT_s]\sim$Unif$(-1,1)$, $T_s=10 ms$ is the sampling interval, and $n$ is the discrete sampling time index.

Other statistical distributions of the bus noise can also be tested, and this particular noise form was selected for the simulation clarity it provides. Without more accurate characterization of power noise, we cannot justify a particular choice of statistical distribution.

\begin{figure}
\centering
\includegraphics[scale=.34]{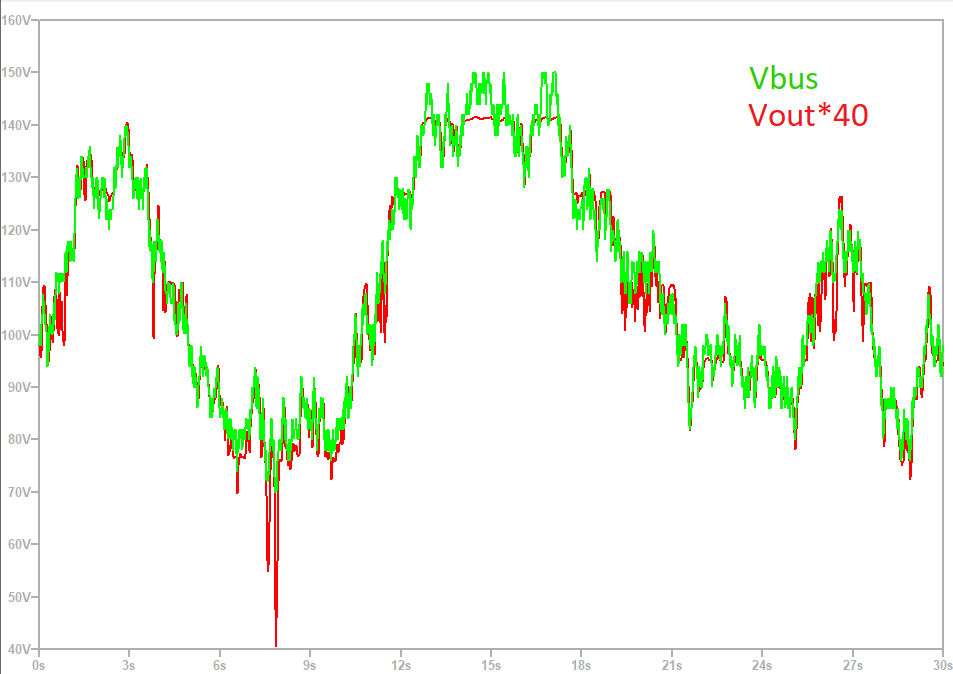}
\caption{A simulated noisy bus voltage is reported across the optical isolation barrier by the AIMQ. Simulated bus voltage is shown in green, and the AIMQ filtered output is shown in dark red (scaled by a factor of 40 to appear on the same plot).}
\label{AIMQsim}
\end{figure}

The AIMQ is able to faithfully follow the bus voltage and provide accurate telemetry, even with only 8 voltage levels, corresponding to an ADC with a bit depth of only 3 bits. Figure \ref{AIMQsim} shows the bus voltage along with the scaled AIMQ output. The green trace is the noisy bus voltage, and the red trace is the filtered AIMQ output scaled by a factor of 40.

\subsection{Simulating the Efficiency of the One-Hot Concept}
\begin{figure}
\centering
\includegraphics[scale=.23]{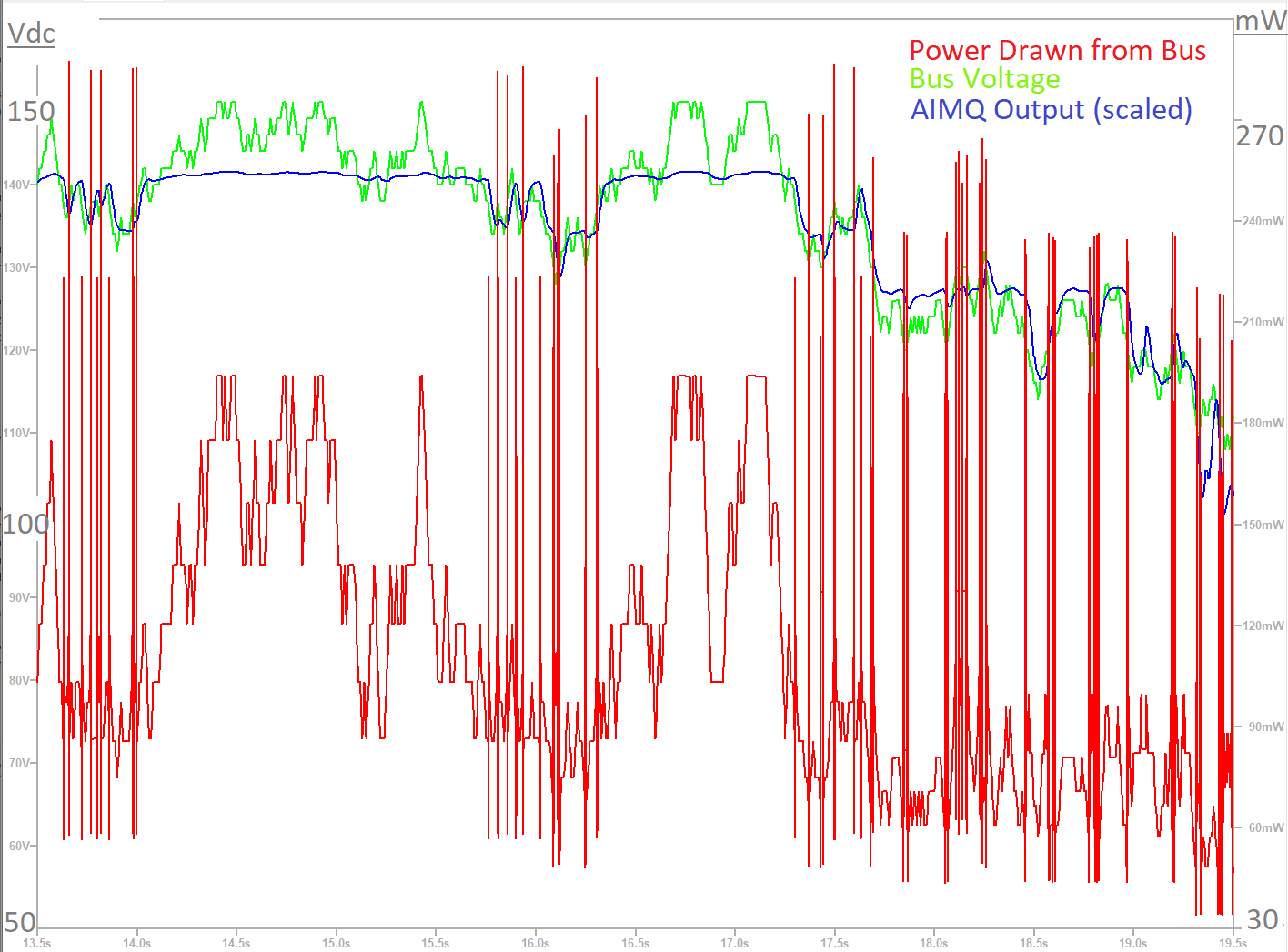}
\caption{Continuous power consumption of the AIMQ. Spikes appear during channel transitions.}
\label{AIMQsim power}
\end{figure}

The efficiency of the AIMQ is a consequence of the one-hot philosophy. In order to follow the bus voltage, it is only necessary for the active channel to conduct enough current to reach the forward current requirement of the input LED of the optocoupler. By ensuring that only one channel conducts at a time, we minimize the total current draw of the AIMQ. However, during the transition between active channels, there will be a current spike as more than one channel is momentarily active. To clean up the AIMQ output waveform, we include an active Sallen-Key low-pass filter. Figure \ref{AIMQsim power} shows the power consumption as the AIMQ monitors the input voltage, with spikes occuring during channel transitions. For a clean input voltage, the AIMQ consumes on the order of 100-200 mW of primary-side power. For reference, an alternative primary-side voltage telemetry solution using an isolation amplifier consumed about 25 mW of primary power through it's voltage divider alone (to provide sufficient bias current so that the measurement voltage divider would be unaffected by the input current of the op-amp), and this comparison also does not include the power consumed by the primary-side 5V/5V DC/DC converter, which is not needed for the AIMQ.

\section{A Complete Model of the AIMQ}
\begin{figure}
\centering
\includegraphics[scale=.2]{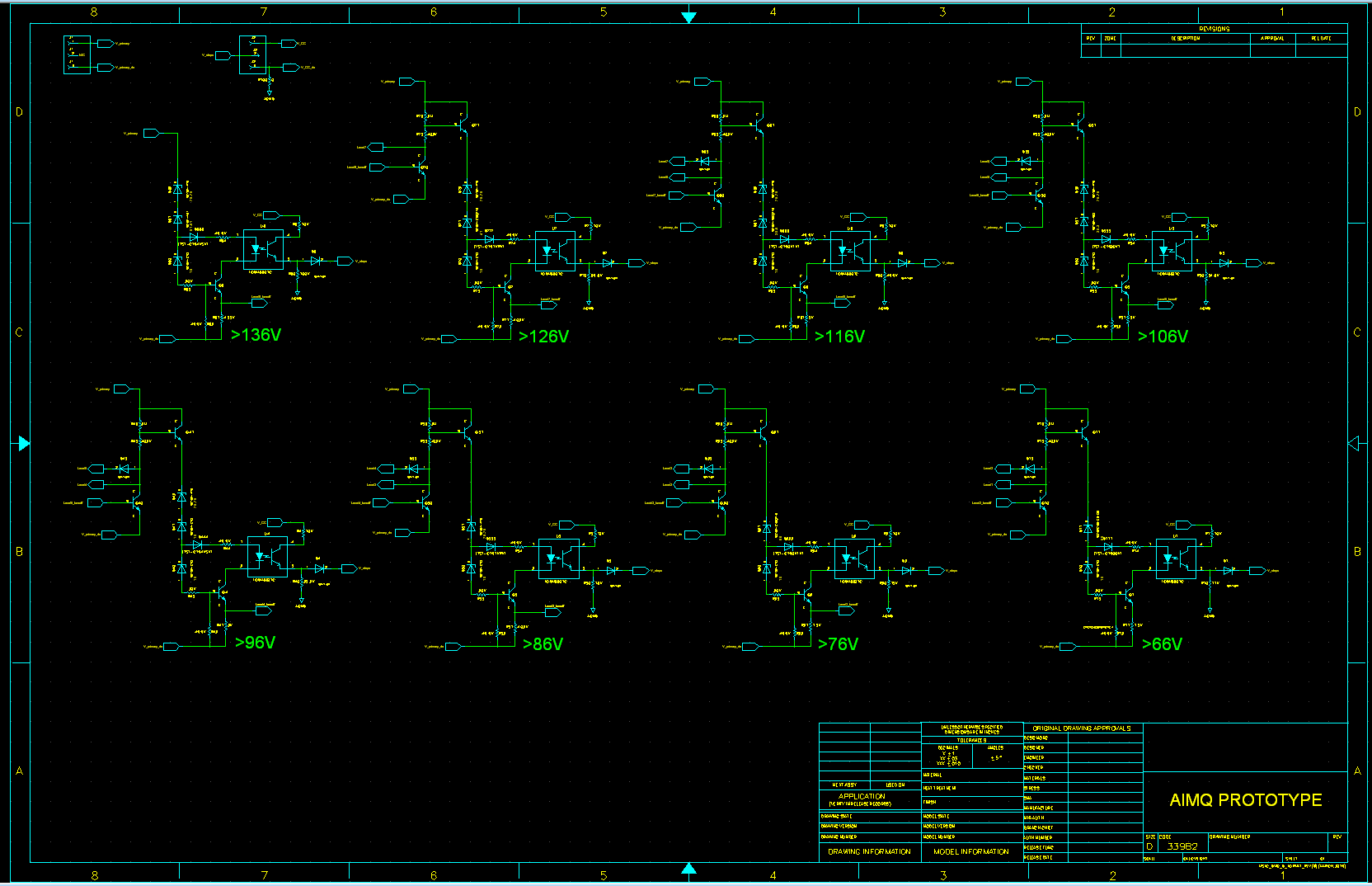}
\caption{8-Level AIMQ prototype schematic.}
\label{AIMQ schematic}
\end{figure}

The simulations shown in Section \ref{AIMQ spice} are from an 8-level model, shown in Figure \ref{AIMQ schematic}. We then developed the full schematic and printed circuit board (PCB) layout using the Mentor Graphics\texttrademark \  Xpedition Suite of PCB design tools \cite{mentor}. We also added colored LEDs for each channel that will illuminate when its channel is active, for demonstration purposes.

\begin{figure}
\centering
\includegraphics[scale=.65, angle=90]{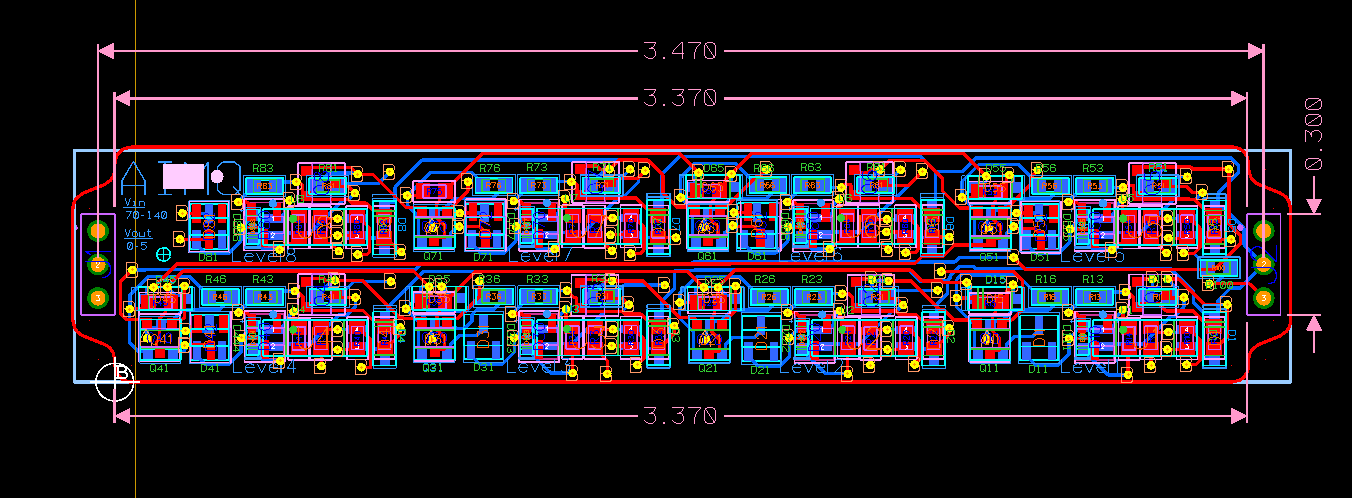}
\caption{8-Level AIMQ prototype board layout.}
\label{AIMQ layout}
\end{figure}

\begin{figure}
\centering
\includegraphics[scale=.72, angle=90]{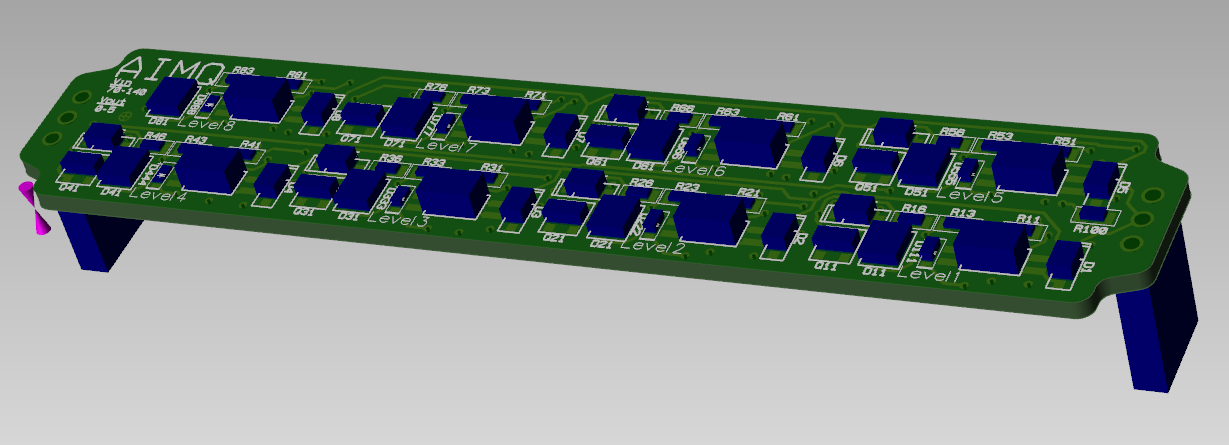}
\caption{3-D isometric view of the AIMQ prototype board.}
\label{3d AIMQ}
\end{figure}

The PCB layout is shown in Figure \ref{AIMQ layout}. The board has 4 layers, and it is densely packed in an effort to show that the device can be a compact replacement for an isolation amplifier with its respective converter and filter. It is designed as a daughterboard that can mate to its parent board via header pins. The top and bottom layers are shown in blue and red, while the inner layers are used as power and ground planes for the two isolated domains, and are not shown in Figure \ref{AIMQ layout}. Figure \ref{3d AIMQ} shows an isometric view of a 3-D rendering of the AIMQ. The female headers on the bottom of the board at both ends mate to a parent board. These 6 pins supply the primary voltage to be measured, the secondary voltage for the optocoupler outputs, and the resulting voltage telemetry signal. One of the 6 pins is not used. The board is roughly the size of a stick of gum, and we are confident that it could be significantly miniaturized, perhaps even reaching a size comparable to the isolation amplifier used in typical designs.

\subsection{AIMQ Parent Board}
\begin{figure}
\centering
\includegraphics[scale=.48]{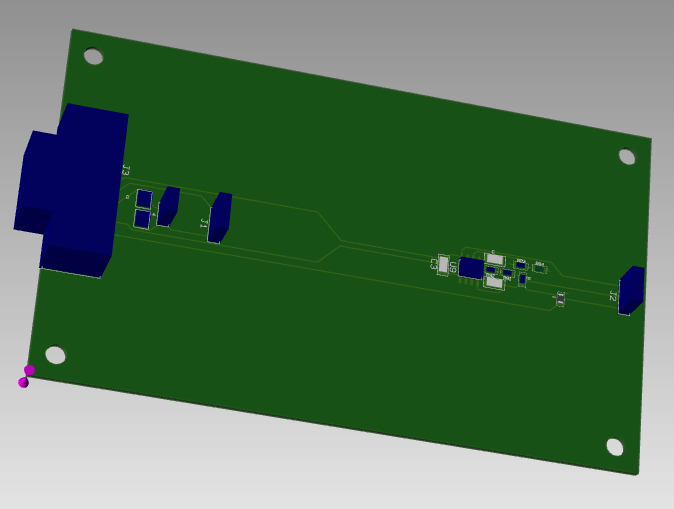}
\caption{3-D view of the AIMQ parent board.}
\label{parent board}
\end{figure}

The AIMQ daughterboard mates to a parent board, shown in Figure \ref{parent board}, via the six header pins. A Sallen-Key active filter  is located on the parent board as well, with a bandwidth selected to smooth the channel transition-induced spikes in the signal, and a gain that can offset any signal loss due to diode drops in the ORing scheme. A 9-pin D-subminiature connector is also included on the parent board for testing.

\subsection{Bandgap References}
Diodes are commonly used as voltage references due to their precision and temperature stability. For a certain Zener voltage it is even possible to achieve a temperature coefficient of $0.001$ \%/K by balancing the Zener breakdown mechanism and the avalanche breakdown mechanism for diode reverse current conduction, because the two voltage-temperature coefficients have slopes with opposite sign \cite{1n3154}. Bandgap references such as the LM113 \cite{lm113} use a shunt regulator topology to achieve tight and adjustable regulation. Achieving such temperature stability from a packaged optically-isolated amplifier or a custom device incorporating highly linear optocouplers marketed for this purpose is problematic due to the inherent time and temperature drift of the optocoupler gain. 

The AIMQ features gain variance immunity, because of the discrete nature of the output signal. The only sources of error are due to the variability of the Zener voltage and the output voltage divider resistances, both of which feature much tighter tolerances than the optocoupler gain. The AIMQ bandwidth and phase margin are dominated by the physics of the optocoupler internal light source and detector, and the present state of the art in linear optocouplers claim frequency responses over 1.5 MHz \cite{IL300}, while high speed digital optocouplers are operating at over 50 Mbps \cite{TLP2367}. The AIMQ is therefore capable of operating at high frequency if necessary. In practice, however, many signals that require isolated measurement do not have high frequency content that is of interest, and the device would more likely be optimized for minimal power by prizing CTR over bandwidth considerations. 

\section{Signal Fidelity}
Fidelity of the transducer used in voltage telemetry may be overemphasized as a requirement, especially in circumstances in which the telemetry multiplexer is polled by the system microprocessor at a low rate, e.g. 1 Hertz. In a case such as this, it is likely that a relatively low-fidelity, 8-level (3-bit) device such as the AIMQ is more than adequate, while being simpler and more efficient than a supposedly higher fidelity alternative. 

\begin{figure}
\centering
\includegraphics[scale=.2]{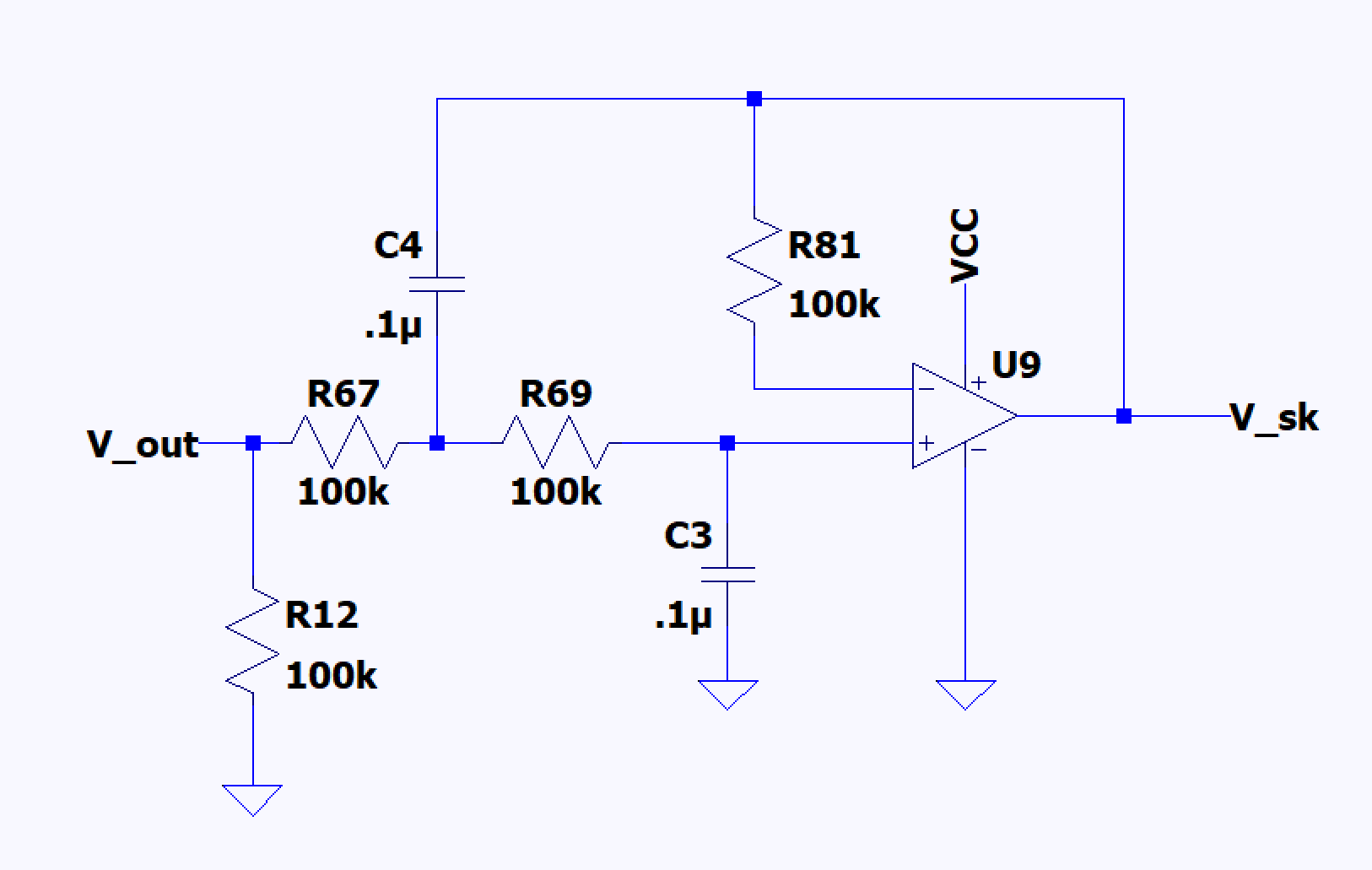}
\caption{A Sallen-Key active filter placed on the AIMQ parent board with crossover frequency of 16 Hz, Gain of 1.1, and Q of 0.5 smooths the AIMQ output signal for telemetry.}
\label{SK filter}
\end{figure}

In addition to the efficiency, EMI, and parts count advantages described earlier, the AIMQ could be packaged as an Integrated Circuit (IC), with user-selectable thresholds determined by placing external Zener diodes. With a sufficient number of parallel quantization optocouplers, the device would provide telemetry data with the required fidelity for many applications.
Furthermore, as optocoupler CTR improves with technology development, the required primary-side power will fall. The required output power is inversely proportional to the load impedance, which will generally be made large
by inclusion of buffer amplifiers. With addition of an emitter follower, the load impedance seen by the AIMQ output can be increased, further reducing power consumption. 

\section{Efficiency Gains of the AIMQ}
The power required by a standard isolated analog linear transducer consists of that consumed by (1) the voltage divider, (2) the supplemental voltage converter, (3) the isolation amplifier's supply current, and (4) the current required to drive the downstream load. The AIMQ is designed to minimize each of these:
First, the current required for a precision Zener diode to enter conduction can be measured in the tens of nanoampere. This is  comparable to the input bias current of a suitable optically isolated amplifier \cite{Ava15}. In
both cases, sufficient current is needed to drive the input LED, and the current must be dissipated across the full input voltage. For this reason, it is desirable to minimize the primary-side current, especially when very high voltages are present.

If the primary side voltage is termed $V_{pri}$ and the isolation amplifier's input bias current is $I_b$, it is common practice (or a rule of thumb) to ensure linear response by providing the voltage divider with $10\cdot I_b$. Total input power is $P_{in}=V_{pri}\cdot 10\cdot I_b$. For the AIMQ, the required input current is determined by the forward current of the optocoupler LED or by the necessary Zener current, whichever is greater, which we shall call $I_{z,f}$. Total AIMQ input power is $V_{pri}\cdot I_{z,f}$. The second and third power terms are linked. The supplemental voltage converter needs to provide the low primary-side voltage for the isolation amplifier, which we shall call $V_{iso}$. The converter has an efficiency which can be termed $\eta$, and the isolation amplifier primary-side supply current we call $I_{pri}$. The total power required is $V_{iso}\cdot I_{pri}/\eta$. 

The AIMQ does not have an equivalent power demand. Furthermore, the supplemental converter requires space, passive components such as capacitors, bias resistors, and filter inductors, and generates noise. In many cases, the converter has a minimum
load current requirement as well. For instance, we have encountered isolated telemetry circuits using converters that call for a minimum load of 10\% of full rated current \cite{CRE1s0505}. Additionally, the isolation amplifier has a secondary-side supply current which is determined by the load impedance. Typically, this would be determined by the ADC or by a buffer amplifier in series with the ADC, with a minimum current stated by the manufacturer under no-load conditions. This current can be termed $I_{sec}$, and the total required power is $P_{sec}=V_{sec}\cdot I_{sec}$. The AIMQ equivalent term is also due to the load impedance, which can be made arbitrarily low through the use of low-power operational amplifier buffer stages. In this case, the minimum AIMQ output power is determined by the output-side voltage divider used to set the discrete signal voltage level, termed $I_{out}$, and the power is $P_{sec,AIMQ}=V_{sec}\cdot I_{out}$. 

Comparing the summed power consumption terms yields the following:

Isolation-amp approach: 
\begin{equation}
   P_{tot1} = V_{pri}\cdot 10 \cdot I_b + V_{iso}\cdot I_{pri}/\eta + V_{sec} \cdot I_{sec} 
\end{equation}

AIMQ: 
\begin{equation}
    P_{tot2} = V_{pri} \cdot I_{z,f} + V_{sec} \cdot I_{out}
\end{equation}

If we assume that the isolation amplifier bias current and the AIMQ Zener/input LED forward current are equal, $I_b=I_{z,f}$, and  assume a worst case DC/DC converter efficiency of $\eta=60\%$, and that the output current of the isolation amplifier is the same as that of the AIMQ $I_{sec}=I_{out}$, we obtain the following ratio of AIMQ power to isolation amplifier power:
\begin{equation}
    \frac{P_{tot2}}{P_{tot1}}=\frac{1}{10+\frac{V_{iso}\cdot I_{pri}/ \eta}{V_{pri}\cdot I_{z,f}}}
\end{equation}
Where we have made the simplifying assumption that $V_{pri}\cdot I_{z,f}>>V_{sec}\cdot I_{out}$. In the case where $V_{pri}=120$ VDC, $V_{iso}=5$ VDC, $I_{pri}=I_{sec}=I_{z,f}=100$ nA, and $\eta=.6$, this comes to:
\begin{equation}
    \frac{P_{tot2}}{P_{tot1}}=\frac{1}{10+\frac{5V\cdot 100nA/.6}{120V\cdot 100nA}}\approx\frac{1}{10.1}
\end{equation}
The AIMQ in this case uses less than 10\% as much power as the isolation amplifier.

\section{Conclusion}
In short, if the supplemental converter efficiency is low, if it has a minimum load requirement, if space is limited, if EMI is a concern, or if optocouplers with high CTR and low forward LED current can be sourced, the AIMQ enjoys significant
advantages over typical isolated measurement approaches.

\bibliographystyle{IEEEtran}

\bibliography{dis}
\end{document}